\documentclass[prd,superscriptaddress,twocolumn, 
preprintnumbers,showpacs,nofootinbib]{revtex4-1} 
\usepackage{amssymb,amsmath,graphicx,epsfig} 
\usepackage{hyperref,color,float,caption,array}
\usepackage{float,url}
\hypersetup{linktocpage=true}
\hypersetup{
colorlinks=true,   
linkcolor=blue, 
citecolor=blue, 
filecolor=blue,    
urlcolor=blue      
}
\def\beq{\begin{equation}} 
\def\eeq{\end{equation}} 
\def\bea{\begin{eqnarray}} 
\def\eea{\end{eqnarray}}

\def\bsmumu{ b \to  s \mu^+ \mu^-}
\def \kstar{K^*}

\def\BKstarmumu{B \to \kstar \mu^+ \mu^-}

\begin{document} 
\title{Explaining the $B \to K^\ast \mu^+ \mu^-$ data with scalar 
interactions}
\author{Alakabha Datta} 
\email{datta@phy.olemiss.edu} 
\affiliation{\normalfont{Department of Physics and Astronomy, 
108 Lewis Hall, University of Mississippi, Oxford, MS 38677-1848, USA}}
\author{Murugeswaran Duraisamy}
\email{duraism@phy.olemiss.edu}
\affiliation{\normalfont{Department of Physics and Astronomy, 
108 Lewis Hall, University of Mississippi, Oxford, MS 38677-1848, USA}}
\author{Diptimoy Ghosh}
\email{diptimoy.ghosh@roma1.infn.it} 
\affiliation{\normalfont{INFN, Sezione di Roma, Piazzale A. Moro 2, 
I-00185 Roma, Italy}}
\begin{abstract} 

Recent LHCb results on the decay $B \to K^\ast \mu^+ \mu^-$ show significant 
deviations from the SM estimates in some of the angular correlations. In 
this paper we study the possibility of explaining these deviations using new 
scalar interactions. We show that new dimension-6 four-quark operators of 
scalar and pseudo-scalar type can successfully account for the discrepancy even 
after being consistent with other experimental measurements. 
We also briefly discuss possible extensions of the Standard Model where these 
operators can be generated.

\end{abstract}
\preprint{UMISS-HEP-2013-08}
\maketitle
\section{Introduction}

The Standard Model (SM) has been extremely successful in explaining all the 
measurements till date in particle-physics experiments. The higgs boson, the 
long awaited last missing piece of the SM, has also been discovered recently 
in the Large Hadron Collider (LHC) experiment 
\cite{Aad:2012tfa,Chatrchyan:2012ufa}. At this moment the main goal of LHC 
will be to look for signals of New Physics (NP) and establish experimentally 
the existence of physics beyond the SM. While direct search experiments are 
extremely important in this endeavor, the flavor physics and other low 
energy experiments will play complimentary roles to the direct search 
experiments in particular, if the NP scale is rather high or do not couple 
significantly to the first two generation of quarks. In fact, deviations from 
the SM expectations at the level of $\sim 2\sigma - 4\sigma$ have already been 
reported in recent years in a few observables involving decays and mixing of 
$B$ mesons 
\cite{Matyja:2007kt,Wei:2009zv,Bona:2009cj,Adachi:2009qg,Bozek:2010xy, 
Abazov:2011yk,Lees:2012xj,Lees:2013uzd,Aaij:2013qta}. On the theoretical side 
also various NP explanations of these deviations have been suggested 
\cite{Datta:2002nr,Hiller:2003js,Baek:2004rp,
Altmannshofer:2008dz,Alok:2009tz,Alok:2010zd,Datta:2010yq, 
Bhattacherjee:2010ju,Alok:2011gv,DescotesGenon:2011yn,Dighe:2011du, 
Bobeth:2011st,Altmannshofer:2011gn,Matias:2012xw,Fajfer:2012vx,Datta:2012qk,
Dighe:2012df,Datta:2012ky,DescotesGenon:2012zf,Duraisamy:2013pia,
Lyon:2013gba,Bhattacharya:2013sga,Descotes-Genon:2013wba,Altmannshofer:2013foa,
Gauld:2013qba,Buras:2013qja,Gauld:2013qja}.

The decays involving $\bsmumu$ transition are particularly interesting as 
they are extremely rare in the SM and many extensions of the SM are capable 
of producing measurable effects beyond the SM. In particular, the three body 
decay $\BKstarmumu$ offers a large number of observables in the kinematic and 
angular distributions of the final state particles and some of these 
distributions have also been argued to be less prone to hadronic uncertainties 
\cite{Altmannshofer:2008dz,Alok:2009tz,Alok:2010zd,Alok:2011gv,
Becirevic:2011bp,Bobeth:2012vn,Matias:2012xw,DescotesGenon:2012zf, 
Descotes-Genon:2013vna}.

The LHCb collaboration has recently measured four angular observables 
($P_4^\prime$, $P_5^\prime$, $P_6^\prime$ and $P_8^\prime$ in the notation of 
\cite{Descotes-Genon:2013vna}) which are largely free from form-factor 
uncertainties, in particular, in the large recoil limit (i.e., low invariant 
mass, $\sqrt{q^2}$, of the di-lepton system). For each of the four observables, 
the data were presented in six $q^2$-bins and quite interestingly, a significant 
deviation of $3.7\sigma$ from the SM expectation was observed only in one of 
the bins ($4.30 < q^2 < 8.68$ GeV$^2$) for only one observable, the 
$P_5^\prime$. 
It is worth mentioning here that there is still considerable amount of 
theoretical uncertainty due to (unknown) power corrections to the 
factorization framework \cite{Jager:2012uw}. Hence, there is a possibility that the observed 
deviation will be resolved once deeper understanding of these corrections 
is achieved. In this paper we take the observed deviation at the face value 
and study its possible explanation from physics beyong the SM.

Note that the observable $P_5^\prime$ is related to the 
observable ${\mathcal S}_5$ defined in 
\cite{Altmannshofer:2008dz,Altmannshofer:2013foa}, see Table.~1 in 
\cite{Altmannshofer:2013foa} for a precise comparison. 
We would like to mention here that the observable ${\mathcal S}_5$ 
is exactly the same (apart from an overall normalization 
factor of 4/3) 
to the Longitudinal-Transverse asymmetry $A_{LT}$ which we defined in our 
earlier work \cite{Alok:2010zd} in the following way,
\bea
A_{LT} = \frac{\int^{\pi/2}_{-\pi/2}d\phi\left\{(\int^1_0 - \int^0_{-1}) 
d\cos {\theta_{K}} \frac{d^3\Gamma}{dq^2d\phi d\cos {\theta_{K}}}\right\}}
{\int^{\pi/2}_{-\pi/2}d\phi\left\{(\int^1_0 + \int^0_{-1}) d\cos {\theta_{K}} 
\frac{d^3\Gamma}{dq^2d\phi d\cos {\theta_{K}}}\right\}}
\eea
where $\theta_{K}$ and $\phi$ are two of the total three angles 
(the other angle $\theta_\mu$ is integrated) in the full angular 
distribution of $B \to K^*(\to K \pi)\mu^+ \mu^-$ (see Fig.~9 
in \cite{Alok:2010zd} for a diagrammatic illustration).

In the SM, the $b \to s$ flavor transition is governed by the 
Effective Hamiltonian,
\beq
{\mathcal H}_{\rm eff}^{SM} = - 
\frac{4 G_F}{\sqrt{2}}\, (V_{ts}^\ast V_{tb}) \,  
\sum_{i=1}^{10} C_i \, {\mathcal O}_i
\eeq
and the decay $B \to K^\ast \mu^+ \mu^-$ proceeds via the three 
operators namely,
\bea 
{\mathcal O}_7 &=& \frac{e}{16\pi^2} \,  
m_b \left (\bar s \sigma_{\alpha \beta} P_R b \right )F^{\alpha \beta} \; , 
\nonumber \\ 
{\mathcal O}_9 &=& \frac{\alpha_{\rm em}}{4 \pi} \,  
\left (\bar s \gamma_\alpha P_L b \right ) (\bar \mu \gamma^\alpha \mu) 
\; \; \rm and \nonumber \\ 
{\mathcal O}_{10} &=& \frac{\alpha_{\rm em}}{4 \pi} \, 
\left (\bar s \gamma_\alpha P_L b \right ) 
(\bar \mu \gamma^\alpha \gamma_5 \mu) 
\eea
with the corresponding Wilson Coefficients $\{C_7, C_9, C_{10}\}$ $\simeq$ 
\{−0.3,4.1, −4.3\} at the scale $\mu=4.8$ GeV. In models beyond the SM new 
chirally flipped ($P_{L(R)}\to P_{R(L)}$) operators ${\mathcal O}_7^\prime$, 
${\mathcal O}_9^\prime$, ${\mathcal O}_{10}^\prime$ may also be generated. It 
was pointed out in \cite{Alok:2010zd} that $A_{LT}$ is particularly sensitive 
to the operators ${\mathcal O}_9$, ${\mathcal O}_9^\prime$, ${\mathcal 
O}_{10}$ and ${\mathcal O}_{10}^\prime$. In fact, a global fit to the NP 
contribution ($\Delta C_{7,9,10}, \Delta C_{7,9,10}^\prime$) to the above six 
Wilson coefficients taking into account the recent LHCb data along with the 
existing data on some other rare and radiative $b \to s$ modes was performed 
in \cite{Descotes-Genon:2013wba} (see also \cite{Beaujean:2013soa}) with the 
conclusion that the deviations seen in the LHCb experiment can be explained 
by just adding a large negative contribution to the Wilson Coefficient $C_9$
\footnote{See however, reference \cite{Hambrock:2013zya} for a possible subtlety.},
\beq
\Delta C_9 \approx -1.5 \; .
\label{sol1}
\eeq
A similar fit to the Wilson Coefficients was also performed in 
\cite{Altmannshofer:2013foa} with a slightly different conclusion. They 
reported the best fit solution to be the one with the presence of NP 
contributions to both $C_9$ and $C_9^\prime$,
\beq
\Delta C_9 \approx -1.0, \; \Delta C_9^\prime \approx 1.0 \; .
\label{sol2}
\eeq
Note that the solutions above are rather unusual as most NP models would in 
general produce not only new contributions to $C_9$ and $C_9^\prime$ but also 
to other operators. In fact, the new $Z^\prime$ boson considered in 
Ref.~\cite{Gauld:2013qba,Gauld:2013qja} to explain the data indeed had rather 
non-standard couplings to the fermions. It is also worth mentioning that 
the scalar or pseudo-scalar operators of the form $(\bar s P_{L(R)} b )
(\bar \mu \mu)$ and $(\bar s P_{L(R)} b )(\bar \mu \gamma_5 \mu)$ 
cannot explain the data owing to their very little effect on 
$A_{LT}$ \cite{Alok:2010zd} in particular, once the consistency with the 
measured branching ratio of $B_s \to \mu^+ \mu-$ is taken into account 
\footnote{In this context, it is also quite interesting to investigate 
the effect of Tensor operators which definitely deserves a separate dedicated 
study and will be presented in a future publication \cite{bksll:future}.}.

In this work we instead consider new four-quark scalar interactions that 
couple the third generation quarks. Possible mixing in the quark sector then 
lead to flavor changing $b \to s$ transitions. Note that there is no direct 
contribution to the decay $\bsmumu$ in this case but it can arise at the one 
loop level. As we will show explicitly in the next sections, in this way we 
can generate new contributions to $C_9$ and $C_9^\prime$ with negligible 
effect on the other operators. In fact, such four-quark scalar operators 
involving third generation of quarks are rather motivated after the discovery 
of the higgs particle and can arise in many extensions of the SM e.g., 
topcolor models \cite{Hill:1994hp,Kominis:1995fj,Buchalla:1995dp}, 
R-parity violating SUSY and multi-higgs models \cite{Datta:1998yq}.

The precise definition of the NP operators will be given in the next section. 
In Sec.~\ref{bsmixing} we will compute the constraints on these operators 
from $\overline{B_s}-B_s$ mixing. Their effect on the $B \to K^\ast \mu^+ \mu^-$ 
decays will be discussed in Sec.~\ref{bksll}. We will stop in 
Sec.~\ref{conclusion} after making some concluding remarks.

\section{New Physics Operators}
\label{np_operators}

As we mentioned in the previous section, in this work we consider effective 
four-quark scalar interactions of the form,
\bea
{\mathcal H}_{\rm eff}^{\rm NP} & = & - \frac{{\mathcal G}_1}{\Lambda^2}
[{\overline s}(1-\gamma^5)b]\,[{\overline b}(1+\gamma^5) b] \nonumber \\
&&-\frac{{\mathcal G}_2}{\Lambda^2}
[{\overline s}(1+\gamma^5)b]\,[{\overline b}(1-\gamma^5) b] + \rm h.c. \
\label{operator}
\eea
which are assumed to be generated by unknown short-distance physics beyond 
the SM. Here $\Lambda$ is the scale of NP and ${\mathcal G}_1$ and 
${\mathcal G}_2$ are the Wilson Coefficients which parameterize our ignorance 
about the underlying microscopic theory. 

In order to proceed with our 
calculations we will not need to work with specific models that can generate 
these operators and hence, we will take Eq.~\ref{operator} as the starting 
point of our phenomenological analysis. However, as an existence proof 
we briefly mention here the topcolor model of ref.~\cite{Hill:1994hp}. 
In such models the top quark participates in a new strong interaction which 
is assumed to be spontaneously broken at some high energy scale $\Lambda$.
The strong interaction, though not confining, leads to the formation of 
top condensate $\langle{\overline t}_Lt_R\rangle$ resulting in scalar 
bound states in the low energy spectrum of the theory which couple strongly 
to the $b$ quark \cite{Kominis:1995fj,Buchalla:1995dp}. 
Integrating out these scalar bound states generates, 
in the weak interaction basis (denoted by $b^\prime$ below), effective four 
fermion operator of the form 
\bea
{\overline b^\prime} \,(1+\gamma_5) \, b^\prime \; {\overline b^\prime}\,(1-\gamma_5)\, b^\prime \; ,
\eea
with possibly rather large couplings \cite{Kominis:1995fj,Buchalla:1995dp}.
The above operator then generates the operators in Eq.~\ref{operator} once 
the quark mass matrices are diagonalized making ${\mathcal G}_{1,2}$ dependent 
also on the mixing matrices of the left and right chiral down type quarks.

\section{$\overline{B_s}-B_s$ mixing}
\label{bsmixing}

\begin{figure}[t!]
\begin{tabular}{c}
\includegraphics[scale=0.4]{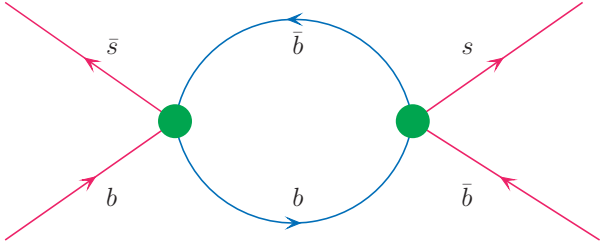}
\end{tabular}
\caption{Feynman diagram showing $\overline{B_s}-B_s$ mixing generated from the 
operators in Eq.~\ref{operator}. \label{fig1}}
\end{figure}

The four-quark operators in Eq.~\ref{operator} will clearly contribute to the 
$\overline{B_s}-B_s$ mixing at the one loop level (See Fig.~\ref{fig1}). 
Taking one operator at a time, the diagram in Fig.~\ref{fig1} will generate the 
following operators, 
\bea
{\mathcal O_1} & = &  {\mathcal K}_1 \; 
[{\overline {s}}(1-\gamma^5)b]\,[{\overline{s}}(1-\gamma^5)b] \; \; 
\rm and \; \nonumber\\
{\mathcal O_2} & = & {\mathcal K}_2 \;
[{\overline {s}}(1+\gamma^5)b]\,[{\overline{s}}(1+\gamma^5)b] \; , 
\eea
where the effective couplings ${\mathcal K}_1$ and ${\mathcal K}_2$ are 
given by, 
\beq
{\mathcal K}_{1(2)} = 
-\frac{3 {{\mathcal G}_{1(2)}^2}}{2 \pi^2 \Lambda^2} 
\frac{m_b^2}{\Lambda^2}
\log \left(\frac{\Lambda^2}{m_b^2}\right) \; . 
\eeq

The magnitude of the NP contribution to the mass difference in 
$\overline{B_s}-B_s$ system can now be written as
\bea
\hspace{-3mm}|\Delta M_{B_s}^{\rm NP}| & = & |{\mathcal K}_{1(2)}| 
\frac{|\langle \overline{B_s^0}|[{\overline {s}}(1 \mp \gamma^5)b]\,
[{\overline{s}}(1 \mp \gamma^5)b]|
{B_s^0}\rangle|}{2 m_{_{B_s}}} \, 
\eea
where $m_{_{B_s}}$ is the mass of the $B_s$ meson. 
With the following definition of the matrix element \cite{Buras:2001ra},  
\bea
&&\langle {\overline{B_s^0}|[{\overline {s}}(1-\gamma^5)b]\,
[{\overline{s}}(1-\gamma^5)]|B_s^0}\rangle  \\ 
&&\phantom{spacespace}=-\frac{5}{3} 
\Bigg(\frac{{m_{B_s}}}{m_b + m_s}\Bigg)^2 {m_{B_s}}^2{f_{B_s}}^2 
{\mathcal B}_{B_s}, \nonumber
\eea
where $f_{B_s}$ and ${\mathcal B}_{B_s}$ are the decay constant and 
relevant bag-parameter respectively, one can now write 
\beq
|\Delta M_{B_s}^{\rm NP}| = 
\frac{5}{6} \Bigg(\frac{{m_{B_s}}}{m_b + m_s}\Bigg)^2 
{m_{B_s}} {f_{B_s}}^2 {\mathcal B}_{B_s}  
|{\mathcal K}_{1(2)}| \, .
\eeq

\begin{figure}[b!]
\begin{tabular}{c}
\includegraphics[scale=0.9]{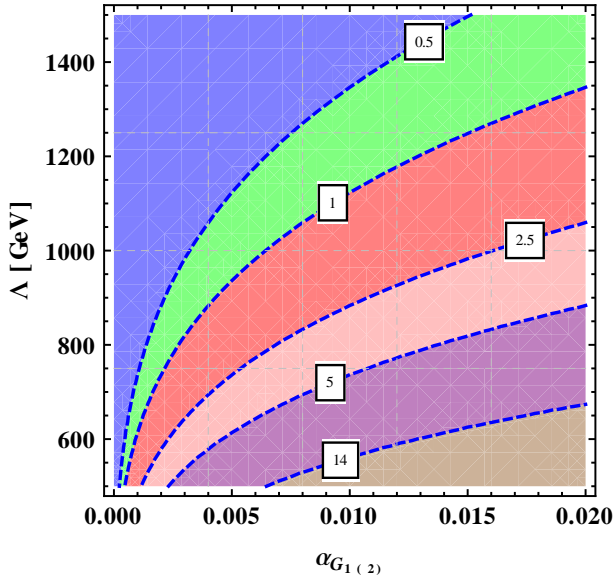}
\end{tabular}
\caption{ Contours of $|\Delta M_{B_s}^{\rm NP}|$ (ps$^{-1}$) in the 
$\alpha_{_{{\mathcal G}_{1(2)}}} - \Lambda$ plane. \label{fig2}}
\end{figure}

In Fig~\ref{fig2} we show the contours of $|\Delta M_{B_s}^{\rm NP}|$ in the 
\mbox{$\alpha_{_{{\mathcal G}_{1(2)}}} - \Lambda$} plane 
($\alpha_{_{{\mathcal G}_{1(2)}}} \equiv {\mathcal G}_{1(2)}^2/4\pi$) 
taking the values of the other parameters to be 
$m_b=4.8$ GeV, 
$m_{B_s}=5.37$ GeV, $f_{B_s} = 225$ MeV  and ${\mathcal B}_{B_s}(m_b) = 0.80$.

The mass difference $\Delta M_{B_s}$ has been very precisely measured with its 
value given by \cite{Amhis:2012bh}, 
\beq
\Delta M_{B_s}^{\rm Exp} =  17.69 \pm 0.08 \; \rm ps^{-1}
\eeq
which is consistent with the SM expectation \cite{Lenz:2011ti}, 
\beq
\Delta M_{B_s}^{\rm SM} =  17.3 \pm 2.6 \; \rm ps^{-1}. 
\eeq
We will conservatively demand that the coupling ${\mathcal G}_{1(2)}$ and 
the NP scale $\Lambda$ satisfy the constraint 
\beq
|\Delta M_{B_s}^{\rm NP}| \lesssim 2.5 \, \rm ps^{-1}. 
\eeq
\section{Contribution to $\bsmumu$}
\label{bksll}

\begin{figure}[h!]
\begin{tabular}{c}
\includegraphics[scale=0.34]{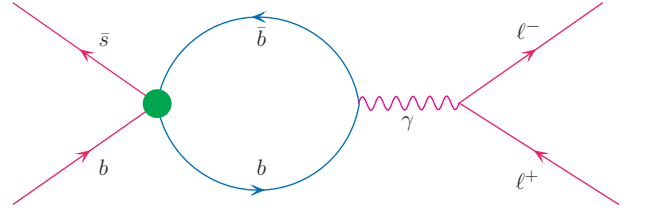}
\end{tabular}
\caption{Feynman diagram showing how the operator in Eq.~\ref{operator} 
contributes to the decay $b \to s l^+l^-$.
\label{fig3}}
\end{figure}

The Effective Hamiltonian ${\mathcal H}_{\rm eff}^{\rm NP}$ of Eq.~\ref{operator} 
generates the effective vertices ${\overline s} b \gamma$, ${\overline s} b g$ and 
${\overline s}b Z$ at the one loop level, as shown in Fig.~\ref{fig3}. The 
vertices with a $\gamma$ or a $Z$ can now contribute to $b \to s l^+l^-$ 
decay once a lepton pair is attached to them. Note that the operators 
${\mathcal O}_7$ or ${\mathcal O}_7^\prime$ are not generated in this way (we 
will see this explicitly below), hence there is no new contribution to the 
decay $b \to s \gamma$.

%
%
A computation of the digram in Fig.~\ref{fig3} (without the lepton pair attached) 
gives the effective vertex for ${\overline s}b\gamma$ to be 
\bea
\label{Rmu}
\includegraphics[scale=0.3]{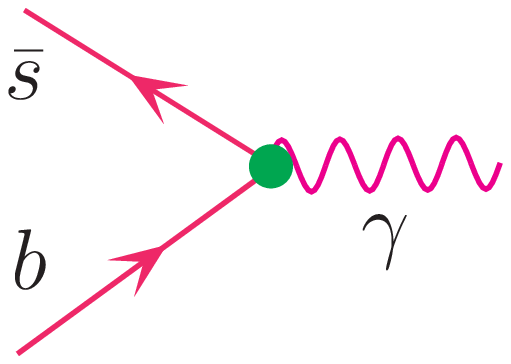} = -\sqrt{4 \pi \alpha_{em}}
\frac{e_b}{\Lambda^2}{\overline s}
\left[{\mathcal G}_1 {\mathcal R}_1^\mu + {\mathcal G}_2 {\mathcal R}_2^\mu \right]b 
\; A_\mu \; \nonumber \\ 
\eea
where
\bea
&&{\mathcal R}_{1(2)}^\mu = \nonumber \\
&&\frac{1}{2\pi^2}\int^1_0 dx \; x(1-x)
\left\{{\rm Ln}\left(\frac{\Lambda^2}{m_b^2}\right)-{\rm Ln}
\left(1 - \frac{q^2}{m_b^2}x(1-x)\right)\right\} \nonumber \\
&&\phantom{space:space}\left[\gamma^{\mu}q^2 - 
q^{\mu} {q\hspace{-1.7mm}\slash}\right] \frac{(1\pm\gamma_5)}{2} \; .
\eea

Here $e_b=-\frac{1}{3}$, the electric charge of the $b$-quark in units of electron 
charge and $q^\mu$ is the 4-momenta of the photon. It is clear from the above 
expression that the amplitude for on-shell photon production is identically 
zero, as claimed in the previous paragraph.

It is now straightforward to calculate the effective vertex for the decay of 
our interest $b \to s l^+ l^-$ by attaching a lepton pair to the virtual photon. 
This gives, 
\bea
\label{bsll}
&&\includegraphics[scale=0.3]{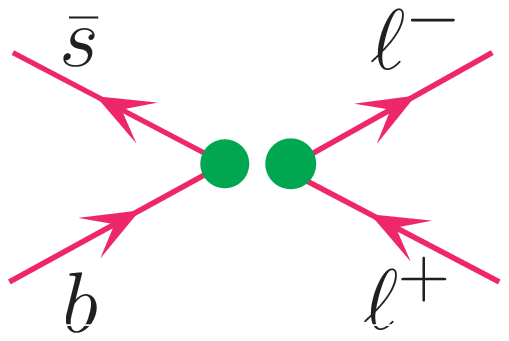} = - (4 \pi \alpha_{em}) 
\frac{e_b}{\Lambda^2} \frac{1}{2 \pi^2} \times \nonumber \\
&&\int^1_0 dx \; x(1-x)
\left\{{\rm Ln}\left(\frac{\Lambda^2}{m_b^2}\right)-{\rm Ln}
\left(1 - \frac{q^2}{m_b^2}x(1-x)\right)\right\}\nonumber \\
&&\left[{\mathcal G}_2 {\mathcal O}_9+   
{\mathcal G}_1 {\mathcal O}_9^\prime\right]
\eea
Note that the the $q^{\mu}$ term in Eq.~\ref{Rmu} does not contribute 
due to electromagnetic gauge invariance. As the contribution coming from a 
$Z$ exchange is suppressed with respect to the $\gamma$ exchange by a factor 
of $q^2/M_Z^2$, we do not include the $Z$ contribution. This also means the 
the new contributions to $C_{10}$ and $C_{10}^\prime$ 
are extremely tiny.

\begin{figure}[h!]
\begin{tabular}{c}
\includegraphics[scale=0.85]{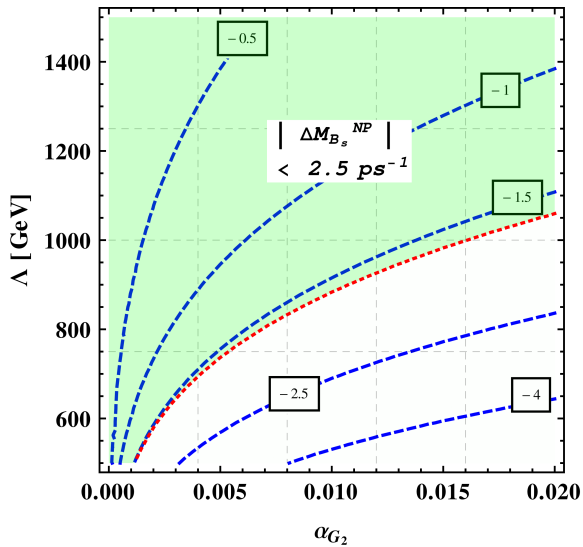}
\end{tabular}
\caption{Contours (blue, dashed) of $\Delta C_9$ in the 
$\alpha_{_{{\mathcal G}_{2}}} - \Lambda$ plane. The green (shaded) region 
above the red (dotted) curve has $|\Delta M_{B_s}^{\rm NP}| < $ 2.5 ps$^{-1}$.
\label{fig4}}
\end{figure}

Comparing Eq.~\ref{bsll} with Eq.~\ref{operator} we can now calculate 
the NP contribution to the Wilson Coefficient $C_9$ and $C_9^\prime$. This reads,
\bea
\label{c9}
&&\Delta C_9 = \frac{2 \sqrt{2} e_b {\mathcal G}_2}{G_F 
\Lambda^2 (V_{ts}^*V_{tb})} \times \nonumber \\
&&\int^1_0 dx \; x(1-x)
\left\{{\rm Ln}\left(\frac{\Lambda^2}{m_b^2}\right)-{\rm Ln}
\left(1 - \frac{q^2}{m_b^2}x(1-x)\right)\right\} \nonumber
\eea
\bea
&& \phantom{C_9}= \frac{2 \sqrt{2} e_b {\mathcal G}_2}{G_F \Lambda^2 (V_{ts}^*V_{tb})} 
\times  \\
&& \left\{\frac{1}{6} {\rm Ln}\left(\frac{\Lambda^2}{m_b^2}\right) - 
\int^1_0 dx \; x(1-x)\, {\rm Ln}
\left(1 - \frac{q^2}{m_b^2}x(1-x)\right)\right\} . \nonumber
\eea

The expression for $\Delta C_9^\prime$ can be obtained from Eq.~\ref{c9} 
after replacing ${\mathcal G}_2$ by ${\mathcal G}_1$. Although $\Delta C_9$ 
is a function of the di-lepton invariant mass $q^2$, the variation in 
$\Delta C_9$ in the whole $q^2$ range is less than 1\% and thus, we will 
neglect this variation below.

In Fig.~\ref{fig4} we show the contours of $\Delta C_9$ in the 
$\alpha_{_{{\mathcal G}_{1(2)}}} - \Lambda$ plane. 
The green shaded region above the red (dotted) contour satisfies 
the constraint $|\Delta M_{B_s}^{\rm NP}| < $ 2.5 ps$^{-1}$. 
Thus, Fig.~\ref{fig4} clearly reveals that the value 
$\Delta C_9 \approx -1.5$ can indeed be achieved keeping the $\overline{B_s}-B_s$ 
mixing completely under control and for reasonable choices of ${\mathcal G}_2$ 
and $\Lambda$. In fact, turning on both the couplings ${\mathcal G}_1$ and 
${\mathcal G}_2$ with opposite sign can even reproduce the solution in 
Eq.~\ref{sol2}.

\section{Conclusion}
\label{conclusion}
%
In this paper we have studied the possibility of explaining certain deviations 
from the SM expectations 
in the angular distribution of the decay $B \to K^\ast \mu^+ \mu^-$ observed 
recently by the LHCb collaboration. We have shown that new dimension-6 four-quark 
operators of scalar and pseudo-scalar type can naturally account for these deviations 
without conflicting with other experimental measurements. This is in contrast 
to generic scalar 4-fermion operators that would in general give rise to new contributions 
to other decays like $b \to s \gamma$, $B_s \to \mu^+ \mu^-$ etc. and hence 
would be very tightly constrained.
We have also briefly mentioned how well known extensions of the SM can generate 
these dimension-6 operators. Detailed phenomenological analysis of these models 
in particular, in view of the large amount of available experimental data, should 
be carried out and will be presented elsewhere. 

\vspace*{1mm}
{\bf Acknowledgments:}
DG is supported by the European Research Council under the European Union's 
Seventh Framework Programme (FP/2007-2013) / ERC Grant Agreement n.279972. 
DG also thanks Satoshi Mishima for discussions. 
The work of AD and MD were financially supported in part by the National Science 
Foundation under Grant No.\ NSF PHY-1068052.


\end{document}